\documentclass[preprint,aps,prc,showpacs,nofootinbib]{revtex4}
\usepackage{epsfig}
\usepackage{graphicx,amsmath}

\newcommand\ba{\begin{eqnarray}}
\newcommand\ea{\end{eqnarray}}

\newcommand{\be}{\begin{equation}}
\newcommand{\ee}{\end{equation}}

\begin{document}

\title{Charge  Form  Factor  and  Cluster  Structure  of $^6$Li
Nucleus}

\author{G.~Z.~Krumova}
\affiliation{\it University of Rousse, 7017 Rousse,
Bulgaria}

\author{E.~Tomasi-Gustafsson}
\affiliation{\it DAPNIA/SPhN, CEA/Saclay, F-91191 Gif-sur-Yvette Cedex,
France}

\author{A.~N.~Antonov }
\affiliation{\it Institute of Nuclear Research and Nuclear Energy, Bulgarian Academy of Sciences, 1784 Sofia, Bulgaria}

\begin{abstract}
The charge form factor of  ${}^6$Li nucleus is considered on the
basis of its cluster structure. The charge density of  ${}^6$Li is
presented as a superposition of two terms. One of them is a folded
density and the second one is a sum of  ${}^4$He and the deuteron
densities. Using the available experimental data for ${}^4$He and
deuteron charge form factors, a good agreement of the calculations
within the suggested scheme is obtained with the experimental data
for the charge form factor of ${}^6$Li, including those in the
region of large transferred momenta.
\end{abstract}

\maketitle
The extensive studies of the nuclei with $\alpha -$cluster structure
have started since the forties and different theoretical models have
been developed till nowadays. Among various $\alpha -$particle
models (APM) that must be noted are, for example, the single APM
(with ready $\alpha -$particles inside the nucleus, e.g.
\cite{1,2,3}), the dynamical APM of point-like $\alpha -$particles
interacting by $\alpha -$$\alpha $ potentials with solving the
Schroedinger or Faddeev equations (e.g.\cite{4}), the microscopic
APM of Brink, Bloch and Margenau (e.g.\cite{5}), and others,
including more recent approaches (e.g.\cite{6}). However, though a
great number of works within the APM have been devoted to the
structure and interactions of such nuclei, many questions remain
open and deserve further work. This concerns even properties of
long-time investigated nuclei, such as the ${}^6$Li nucleus. It is
known that the structure of the $^6$Li nucleus has some
peculiarities compared to the other 1p-shell nuclei (see e.g.
\cite{6,7,8,9,10,11,12}). The elastic electron scattering data
\cite{13} on the charge form factor and the rms radius of $^6$Li
cannot be explained in the framework of the shell-model by means of
an oscillator parameter $\hbar\omega=15\div16 \ MeV$, the latter
providing a good description of these data for the other 1p-shell
nuclei. The usage of another value of $\hbar\omega$, the same for
the s- and p- nucleons, as well as of two different oscillator
parameters for the s- and p-shells is also not successful. The
situation is similar in the case of the inelastic form factors. The
wave functions of the low-lying states of ${}^6$Li are significantly
different from the commonly accepted and used shell model wave
functions. This fact is important for the analysis of the $(p,2p)$,
$(p,pd)$, $(p,p\alpha)$ reactions on ${}^6$Li, the photonuclear
reactions, the ($^6$Li, d), ($^6$Li, $\alpha)$ reactions and others.

It has been estimated that $^6$Li has a well pronounced cluster
structure and is considered generally as a system consisting of
$\alpha -$ and deuteron clusters in a mutual motion exchanging
nucleons. The small value of the decay threshold
$^6$Li$\to\alpha+d$, the large nuclear radius, etc. give evidence
that the $\alpha -$ and $d -$ clusters in $^6$Li are quite isolated.
In another of the cluster models, the Model of Nucleon Associations
(MNA) (e.g. \cite{12}), the problem of the role of the exchange has
been studied by analyzing the elastic and inelastic form factors of
the Coulomb electron scattering. The antisymmetrization effect turns
out to be substantial only at large values of the isolation
parameter $x\approx 1$, where $x = b/a$ is the ratio between the
relative motion function parameter $b$ and the $\alpha$-particle
function parameter $a$. At the real value $x=0.3\div 0.4$, the
exchange effects are already of no importance. In MNA the value $x =
1$ corresponds to the shell-model structure of $^6$Li, while $x = 0$
corresponds to the cluster model ($\alpha - d$ structure). It has
been found that the isolation parameter $x$ has different values for
nuclei with cluster structure. For instance, $x=0.5\div 0.6$ for
$^9$Be, $x=0.7$ for $^{12}$C and $x=0.8$  for $^{16}$O \cite{14,15}.
The elastic scattering charge form factor, although being sensitive
to the value of $x$, can be described by different models due to the
fact that it is obtained on the base of the charge density
distribution that is averaged over the angular variables. The form
factor of the inelastic quadrupole scattering, however, strongly
depends on $x$ and cannot be described within the shell model. The
MNA provides a good rms radius of $^6$Li \cite{16} and with the
above values of the isolation parameter ($x=0.3\div 0.4$) allows a
proper simultaneous description of the electron elastic and
inelastic scattering but only up to transferred momentum values
$q\sim 2 \ fm{}^{-1}$.

 The aim of the present work is to suggest an
approach in which the $\alpha -d$ cluster structure of $^6$Li to be
checked by calculations of the charge density and the corresponding
charge form factor. We construct a scheme in which the charge
densities of $^4$He and the deuteron are included and the available
experimental data for them can be used to calculate the $^6$Li
charge density, the charge form factor and the latter to be compared
with the experiment. In this sense, our work has a meaning of a
'theoretical experiment' to check the particular cluster structure
of $^6$Li by a comparison of the results of two different
suggestions with the empirical data for this nucleus.

In Section~\ref{sec2} the theoretical scheme, the results of the
calculations and a discussion are presented. The conclusions are
given in Section~\ref{sec3}.

\section{Charge Density and Form Factor of $^6$Li in Relation to
Those of $^4$He and Deuteron}\label{sec2}
Considering the problems
in the description of the $^6$Li charge density and form factor
briefly mentioned above, we made an attempt to study these
quantities on the base of the corresponding ones for $^4$He and the
deuteron within the framework of the $\alpha-d$ cluster structure of
$^6$Li nucleus.

Our first attempt was to describe the charge density of $^6$Li
within the framework of the often used folding procedure. In our
case it is  a folding of the charge densities of $^4$He and the
deuteron:

\begin{equation}
  {\rho}_{\scriptscriptstyle ^6\!Li}^{\scriptscriptstyle ch}\left(\vec{r}\right)=\frac32 \int
    \mathrm{d}\vec{r}\, ' \rho_{\scriptscriptstyle ^4\!He}^{\scriptscriptstyle ch}\left(\vec{r}-\vec{r}\, '\right)
    \rho_{\scriptscriptstyle d}^{\scriptscriptstyle ch}\left(\vec{r}\, '\right)\ . \label{eq:rhofold}
\end{equation}
The charge densities in Eq. (\ref{eq:rhofold}) are normalized to
the number of protons $Z$ ($Z=3$, $2$ and $1$ for $^6$Li, $^4$He
and the deuteron, correspondingly). Substituting $^6$Li charge
density Eq. (\ref{eq:rhofold}) in the definition of the charge
form factor
\begin{equation}
F^{\scriptscriptstyle ch} \left(\vec{q}\,\right)=\frac{1}{Z}\int
     \mathrm{d}\vec{r}\  e^{i\vec{q}.\vec{r}} {\rho}^{\scriptscriptstyle ch}\left(\vec{r}\, \right)
\label{eq:ffdef}
\end{equation}
we obtain
\begin{equation}
F_{\scriptscriptstyle ^6\!Li}^{\scriptscriptstyle
ch}(q)=F_{\scriptscriptstyle ^4\!He}^{\scriptscriptstyle
 ch}(q)\,
    F_{\scriptscriptstyle d}^{\scriptscriptstyle ch}(q)\, e^{\, q^2/(4A^{2/3})}\ ,
 \label{eq:fffold}
\end{equation}
in which the exponential factor approximately accounts for the
centre-of-mass (c.m.) corrections according to \cite{17}.

 In our calculations of the charge form factor of $^6$Li (Eq. (\ref{eq:fffold}))
we use the available experimental data for the charge form factor
of $^4$He (see e.g. \cite{18} and references therein), as well as
the experimental data for the charge form factor of the deuteron.
The latter are those from the Thomas Jefferson Laboratory
experiments in which the deuteron charge form factor was measured
for a first time to a transferred momentum value up to $q=6.64 \,
fm^{-1}$ and the node of the form factor was observed (Abbott et
al. \cite{19,20}). In our calculations for the deuteron charge
form factor we use a best fit parametrization obtained in
\cite{21}. It is represented by Eq. (\ref{eq:ffd1}) - Eq.
(\ref{eq:ffd5}) \cite{21}:
\begin{equation}
F_{\scriptscriptstyle d}^{\scriptscriptstyle
ch}(q^2)=g(q^2)\overline{F}_{\scriptscriptstyle
d}^{\scriptscriptstyle ch}(q^2)\, ,\label{eq:ffd1}
\end{equation}

\begin{equation}
{\overline{F}}_{\scriptscriptstyle d}^{\scriptscriptstyle
    ch}(q^2)=1-\alpha-\beta+\alpha\frac{m_{\omega}^2}{m_{\omega}^2+q^2}+\beta\frac{m_{\Phi}^2}{m_{\Phi}^2+q^2}\,
    ,
\label{eq:ffd2}
\end{equation}
where $m_{\omega}$ and $m_{\Phi}$ are the meson masses ($m_{\omega}=
0.784\   GeV$ and $m_{\Phi}= 1.019\    GeV$). For any values of the
two real parameters $\alpha$ and $\beta$
\begin{equation}
F_{\scriptstyle d}^{\scriptstyle ch}(0)=1\, . \label{eq:ffd3}
\end{equation}
The factor $g$ in  Eq. (\ref{eq:ffd1}) has the form
\begin{equation}
g(q^2)=\frac{1}{(1+\gamma q^2)^{\delta}} \label{eq:ffd4}
\end{equation}
and $\gamma$ and $\delta$ are also real parameters.

The requirement of a node  for $q_0^2 \approx 0.7\  GeV^2$ gives the
following relation between the parameters $\alpha$ and $\beta$:
\begin{equation}
\alpha=\frac{m_{\omega}^2+q_{0}^2}{
q_{0}^2}-\beta\frac{m_{\omega}^2+q_{0}^2}{ m_{\Phi}^2+q_{0}^2}\, .
\label{eq:ffd5}
\end{equation}

The values of two sets of the parameters $\alpha$, $\beta$, $\gamma$
and $\delta$ obtained in \cite{21} by a best fit to the experimental
data, which are used in the calculations of the present work, are
given in Table 1.

\begin{table}[htbp]\centering
\caption{The values of the parameters $\alpha$, $\beta$, $\gamma$
and $\delta$ in the parametrizations I and II, obtained from the
global best fit in \protect\cite{21} (the values of $\alpha$ are
derived from Eq. (\protect\ref{eq:ffd5})).}
{\begin{tabular}{|c|c|c|c|c|}  \hline
 Set &  $\alpha$ & $\beta$ & $\gamma$ & $\delta$ \\ \hline
I     & $5.75\pm 0.07$ & $-5.11\pm 0.09$ & $12.1\pm 0.5$ & $1.04\pm
0.03$ \\ \hline II    & $5.50\pm 0.06$ & $-4.78\pm 0.08$ & $12.1\pm
0.5$ & $1.05\pm 0.03$ \\ \hline
\end{tabular}}
\end{table}

Recent Large-Scale Shell-Model (LSSM) calculations of \cite{22} and
the analysis of the elastic and inelastic electron and proton
scattering data from $^{6,7}$Li have proved the 'clustering'
behavior of these systems. In our work \cite{23} proton, neutron,
charge, and matter densities of a wide range of exotic nuclei
obtained in the Hartree-Fock-Bogolyubov method and in the LSSM (for
He and Li isotopes) have been used for Plane-Wave Born Approximation
(PWBA) and Distorted-Wave Born Approximation (DWBA) calculations of
the related form factors. This makes it possible to analyze the
influence of the increasing number of neutrons on the proton and
charge distributions in a given isotopic chain. The obtained in
\cite{23} theoretical predictions for the charge form factors of
exotic nuclei are a challenge to their measurements in the future
experiments on the electron-radioactive beam colliders in GSI and
RIKEN in order to get detailed information on the charge
distributions of such nuclei.

\begin{figure}[th ]\centering
\includegraphics[width=12cm]{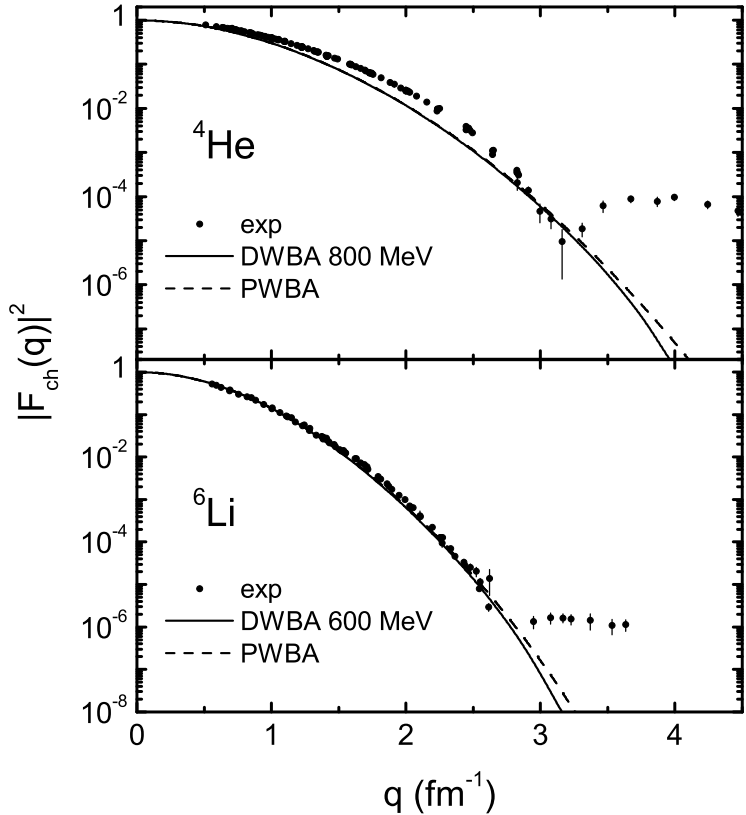}\\
\caption{Charge form factors of the stable isotopes $^4$He and
$^6$Li obtained in \protect\cite{23} using LSSM densities in PWBA
and in DWBA calculations in comparison with the experimental data
\protect\cite{18,24,25,26,27,28,29,30}.}\label{fig1}
\end{figure}

The available experimental data \cite{18,24,25,26,27,28,29,30}for
$^4$He and $^6$Li charge form factors are presented in Fig.
\ref{fig1} in comparison with the results of our PWBA and DWBA
calculations from \cite{23}. One can see a good agreement with the
data up to $q\sim 3\, fm^{-1}$.

\begin{figure}[th ]\centering
\includegraphics[width=10cm]{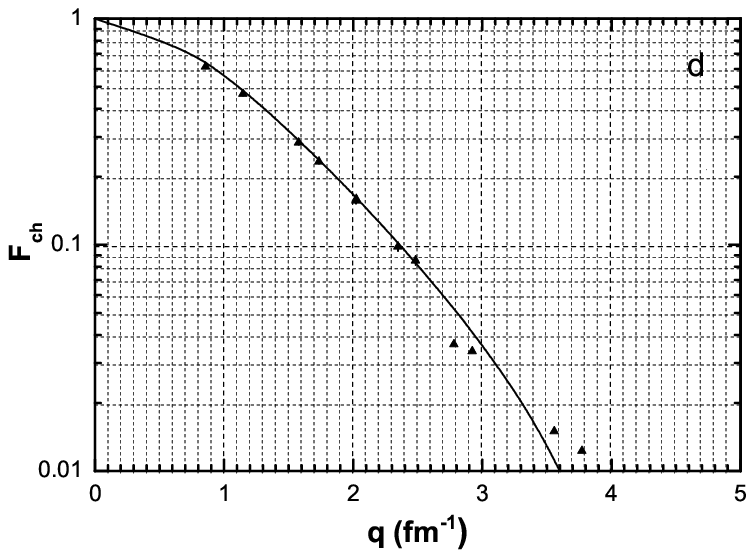}\\
\caption{The charge form factor of the deuteron calculated using
two sets of parameters $\alpha$, $\beta$, $\gamma$ and $\delta$
(with values given in Table 1) and compared with the experimental
data \protect\cite{19,20} up to $q\approx 3.8\, fm^{-1}$.}
\label{fig2}
\end{figure}

In Fig. \ref{fig2} are presented the experimental data for the
deuteron charge form factor \cite{19,20} and the result of the
parametrization from \cite{21} up to $q\approx 3.8\, fm^{-1}$ (with
parameter sets I and II from Table 1).

\begin{figure}[th ]\centering
\includegraphics[width=10cm]{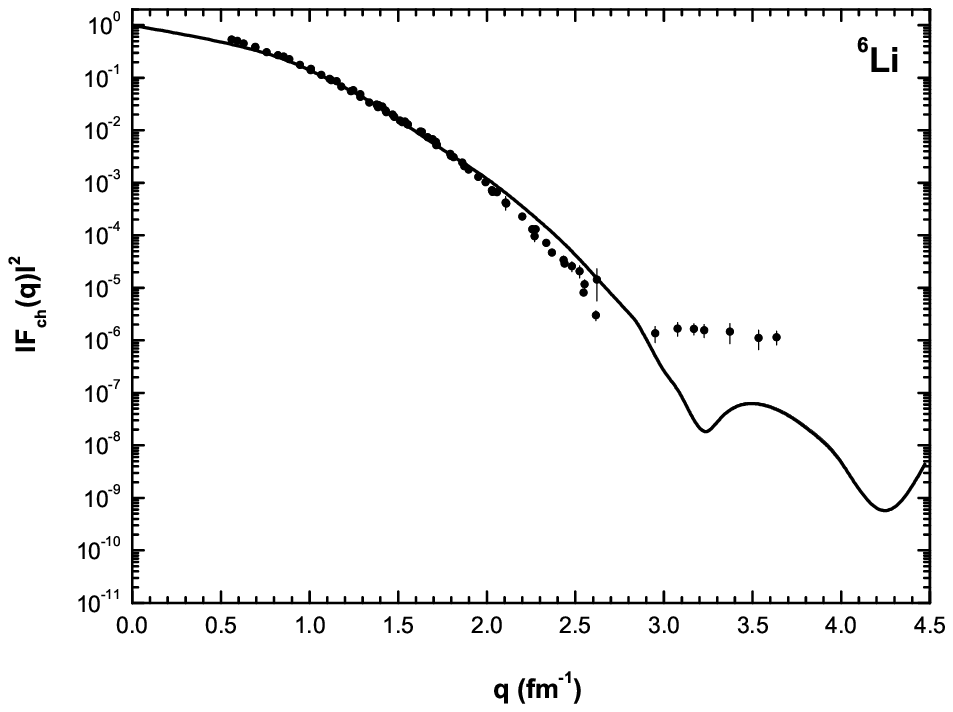}\\
\caption{The charge form factor of $^6$Li calculated by using Eq.
(\protect\ref{eq:fffold}) and the experimental data for the charge
form factors of $^4$He and the deuteron in comparison with the
experimental data (\protect\cite{18,24,25,26,29,30}).}
\label{fig3}
\end{figure}

In Fig. \ref{fig3} are given our results for the squared charge
form factor of $^6$Li calculated by using of Eq. (\ref{eq:fffold})
(taking account of the c.m. correction) and the experimental data
for the charge form factors of $^4$He and the deuteron. For the
latter we used the same parametrization from \cite{21} Eq.
(\ref{eq:ffd1}) - Eq. (\ref{eq:ffd5}) with two sets of parameters
I and II from Table 1. A good agreement with the experimental data
in the interval of transferred momentum $0<q\leq 2.7\, fm^{-1}$
can be seen and a disagreement with the values of the form factor
for larger $q$'s that are related to small values of $r$'s, i.e.
to the central part of the nuclear density. In other words, the
central density can be different from the assumption for the
folding density (Eq. (\ref{eq:rhofold})). We note the similarity
of the results (compared with the data) of the calculated charge
form factor of $^6$Li  for $q\lesssim 2.7\, fm^{-1}$ from the
present work (shown in Fig. \ref{fig3}) with those from \cite{23}
(shown in the down panel of Fig. \ref{fig1}).

The results shown in Fig. \ref{fig3} were the reason to look for
an extension of the approach. Our second suggestion was to
consider the charge density of $^6$Li as a superposition of a
folding term and a sum of the charge densities of $^4$He and the
deuteron with weight coefficients $c_1$ and $c_2$:
\begin{equation}
{\rho}_{\scriptscriptstyle ^6\!Li}^{\scriptscriptstyle
ch}\left(\vec{r}\,\right)=\frac32 c_1\int
    \mathrm{d}\vec{r}\, ' \rho_{\scriptscriptstyle ^4\!He}^{\scriptscriptstyle ch}\left(\vec{r}-\vec{r}\, '\right)
    \rho_{\scriptscriptstyle d}^{\scriptscriptstyle ch}\left(\vec{r}\, '\right)+c_2\left[\,\rho_{\scriptscriptstyle ^4\!He}^
    {\scriptscriptstyle ch}\left(\vec{r}\,\right)+
    \rho_{\scriptscriptstyle d}^{\scriptscriptstyle ch}\left(\vec{r}\, \right)\right]\, .
\label{eq:rhoII}
\end{equation}
The normalization of the densities in Eq. (\ref{eq:rhoII}) to $Z$
leads to the condition for the coefficients
\begin{equation}
c_1 + c_2=1 . \label{eq:c1+c2}
\end{equation}

Using the charge density (Eq. (\ref{eq:rhoII})), the following
expression for the charge form factor of $^6$Li (with the account
for the c.m. correction) is obtained:
\begin{equation}
F_{\scriptscriptstyle ^6\!Li}^{\scriptscriptstyle ch}(q)=\left\{c_1
F_{\scriptscriptstyle ^4\!He}^{\scriptscriptstyle
 ch}(q)\,
    F_{\scriptscriptstyle d}^{\scriptscriptstyle ch}(q)+\frac{c_2}{3}\left[2F_{\scriptscriptstyle ^4\!He}^{\scriptscriptstyle
 ch}(q)+F_{\scriptscriptstyle d}^{\scriptscriptstyle ch}(q)\right]\right\}\, e^{\, q^2/(4A^{2/3})}\, . \label{eq:ffII}
\end{equation}
For $q=0$
\begin{equation}
F_{\scriptscriptstyle ^6\!Li}^{\scriptscriptstyle ch}(0)=1\ .\label{eq:ffII=1}
\end{equation}
The squared charge form factor can be written as:
\begin{equation}
{\mid F_{\scriptscriptstyle ^6\!Li}^{\scriptscriptstyle ch}(q)\mid
}^2=A+B+C\, ,\label{eq:ffII^2}
\end{equation}
where $A$, $B$ and $C$ represent the contributions to the charge
density of $^6$Li of the folding term ($A$), of the sum of the
charge densities of $^4$He and the deuteron ($B$) and the
interference term ($C$). Their explicit expressions are:
\begin{equation}
A={c_1}^2{\mid F_{\scriptscriptstyle ^4\!He}^{\scriptscriptstyle
 ch}(q)\mid}^2 {\mid F_{\scriptscriptstyle d}^{\scriptscriptstyle ch}(q)\mid}^2\, e^{\, q^2/(2A^{2/3})}\, , \label{eq:A}
\end{equation}
\begin{equation}
B=\frac{{c_2}^2}{9}[\, 4\, {\mid F_{\scriptscriptstyle
^4\!He}^{\scriptscriptstyle
 ch}(q)\mid}^2+{\mid F_{\scriptscriptstyle d}^{\scriptscriptstyle ch}(q)\mid}^2
 +4\mid F_{\scriptscriptstyle4\!He}^{\scriptscriptstyle ch}(q)\mid\mid F_{\scriptscriptstyle d}^{\scriptscriptstyle
 ch}(q)\mid]\, e^{\, q^2/(2A^{2/3})}\, , \label{eq:B}
\end{equation}
\begin{equation}
C=\frac23 c_1c_2\mid F_{\scriptscriptstyle
^4\!He}^{\scriptscriptstyle ch}(q)\mid \mid F_{\scriptscriptstyle
d}^{\scriptscriptstyle ch}(q)\mid [\, 2\, \mid F_{\scriptscriptstyle
^4\!He}^{\scriptscriptstyle ch}(q)\mid + \mid F_{\scriptscriptstyle
d}^{\scriptscriptstyle ch}(q)\mid ]\, e^{\, q^2/(2A^{2/3})}\, .
\label{eq:C}
\end{equation}

In the following  Fig. \ref{fig4} are presented the results for
the squared charge form factor of $^6$Li calculated using Eq.
(\ref{eq:ffII^2}) - Eq. (\ref{eq:C}) and the experimental data for
the charge form factor of $^4$He, of the deuteron and with
different sets of the values of the weight coefficients $c_1$ and
$c_2$. The fit of
 Eq. (\ref{eq:ffII^2}) to the experimental data reveals an interval of values of
$c_1=0.975\div0.985$ and, correspondingly, of $c_2=0.025\div0.015$,
for which the results reasonably agree with the experimental data
\cite{18,24,25,26,29,30} within the limits of the experimental
errors. In our opinion, a more reasonable result is obtained for the
case with $c_1=0.979$ and $c_2=0.021$. For the latter we also show
in Fig. \ref{fig5} the contributions of the three terms $A$, $B$ and
$C$.

\begin{figure}[th ]\centering
\includegraphics[width=10cm]{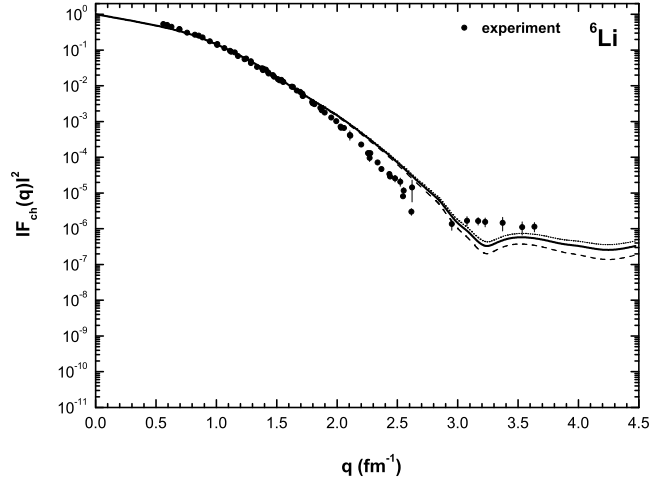}\\
\caption{The charge form factor of $^6$Li (Eqs.
(\protect\ref{eq:ffII}) - (\protect\ref{eq:C})) calculated for
$c_1=0.975$, $c_2=0.025$ (dotted line), $c_1=0.979$, $c_2=0.021$
(solid line), and $c_1=0.985$, $c_2=0.015$ (dashed line). The
experimental data are taken from
\protect\cite{18,24,25,26,29,30}.} \label{fig4}
\end{figure}

\begin{figure}[th ]\centering
\includegraphics[width=10cm]{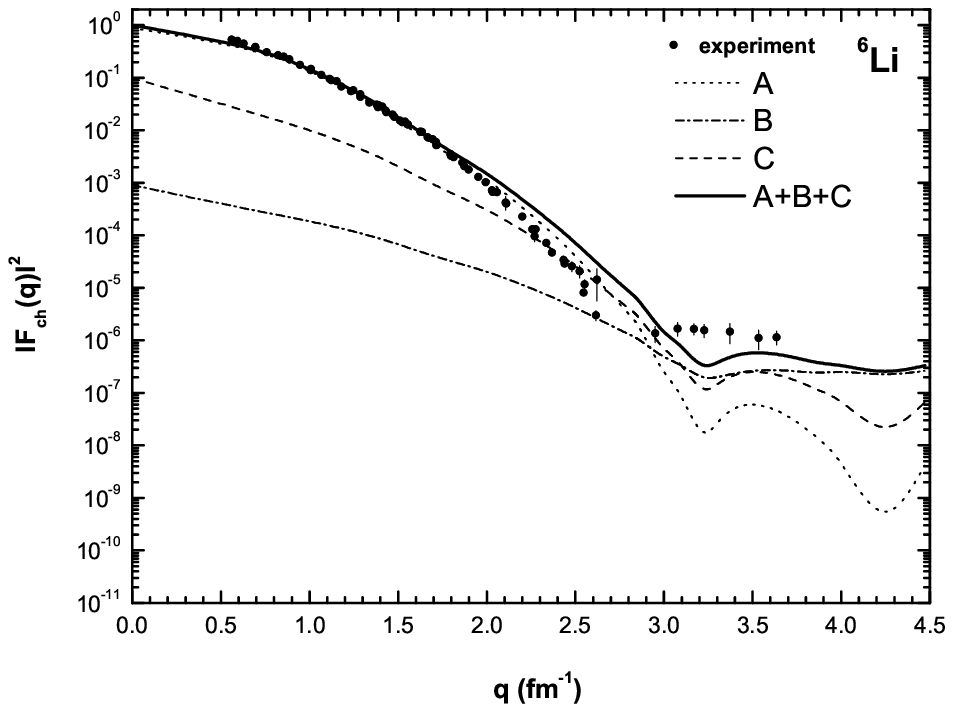}\\
\caption{The same as in Fig. \protect\ref{fig4} for $c_1=0.979$,
$c_2=0.021$. The contributions of $A$, $B$ and $C$ terms are
presented.} \label{fig5}
\end{figure}

The results presented in Fig. \ref{fig4} and Fig. \ref{fig5} prove
that the contribution of the folding density to the charge density
of $^6$Li is about $97.5\div98.5\%$. This corresponds to the
weight of the contribution of the sum of $^4$He and the deuteron
densities of about $2.5\div1.5\%$. It is seen that the term $A$
(Eq. (\ref{eq:A})) describes well the squared charge form factor
of $^6$Li in the interval $0<q\lesssim  2.7\, fm^{-1}$, while the
shell-model cluster density (related to the term $B$, Eq.
(\ref{eq:B}))  is important for the description of the charge form
factor of $^6$Li for the large values of $q$ ($q\gtrsim 3\,
fm^{-1}$), related to the central nuclear density. The
interference term $C$ (Eq. (\ref{eq:C})) has a contribution to the
charge form factor of $^6$Li for $q\gtrsim 3\, fm^{-1}$. The
increase of $c_1$ within the above interval leads to a better
description of the data for $q=1.8\div2.9\, fm^{-1}$, but at the
same time to a decrease of the values of the squared $^6$Li charge
form factor for $q\gtrsim 3\, fm^{-1}$, underestimating the data.

As known, the value of the obtained rms radius is a test for the
consistency of any approach to the description of the nuclear system
structure. The charge rms radius of $^6$Li is given by the
expression:
\begin{equation}
\langle r_{\scriptscriptstyle ^6\!Li}^{\scriptscriptstyle 2}\rangle
=\frac13\int\mathrm{d}\vec{r}\, r^2 {\rho}_{\scriptscriptstyle
^6\!Li}^{\scriptscriptstyle ch}\left(\vec{r}\right)\, .
\label{eq:rms1}
\end{equation}
Substituting the expression for the charge density of $^6$Li (Eq.
({\ref{eq:rhoII})) in Eq. (\ref{eq:rms1}), we obtain:
\begin{equation}
\left\langle r_{\scriptscriptstyle ^6\!Li}^{\scriptscriptstyle
2}\right\rangle = c_1\left[\left\langle r_{\scriptscriptstyle
^4\!He}^{\scriptscriptstyle 2}\right\rangle + \left\langle
r_{\scriptscriptstyle d}^{\scriptscriptstyle 2}\right\rangle\right]
+ \frac{c_2}{3}\left[2 \left\langle r_{\scriptscriptstyle
^4\!He}^{\scriptscriptstyle 2}\right\rangle
 + \left\langle r_{\scriptscriptstyle d}^{\scriptscriptstyle 2}\right\rangle \right]\, . \label{eq:rms2}
\end{equation}
The usage of the experimental data for the rms radii of $^4$He and
the deuteron \cite{25,31}: $${\left\langle
r_{\scriptscriptstyle^4\!He}^{\scriptscriptstyle
2}\right\rangle}^{1/2} =1.676(8)\, fm\, ,$$ $${\langle
r_{\scriptscriptstyle d}^{\scriptscriptstyle
 2}\rangle}^{1/2} =2.116(6)\, fm$$
in Eq. (\ref{eq:rms2}) (with $c_1=0.979$ and $c_2=0.021$) leads to
the following value for the $^6$Li charge rms radius:
 $$\langle r_{\scriptscriptstyle ^6\!Li}^{\scriptscriptstyle
 2}\rangle^{1/2}
=2.684\, fm\, ,$$
 which is in accordance with the experimental
estimations for the charge rms radius of $^6$Li \cite{25,31}:
$$\langle r_{\scriptscriptstyle ^6\!Li}^{\scriptscriptstyle
2}\rangle^{1/2} =2.57(10)\ fm.$$  This could be expected due to
the use of the experimental charge densities of the deuteron and
$^4$He, being combined in a realistic theoretical scheme that
gives a good agreement with the experimental data for the charge
form factor of $^6$Li.

\section[]{Conclusions}\label{sec3}

In the present work we suggest a theoretical scheme for calculations
of the charge density distribution and form factor of $^6$Li in the
framework of the $\alpha-d$ cluster model of this nucleus. The
obtained results can be summarized as follows:

\begin{itemize}
\item Our calculations show a reasonable description
of the charge form factor of $^6$Li on the basis of a
superposition of two density distributions:
\begin{description}
\item [(a)] a folding density obtained from $^4$He and the
deuteron charge densities, and \item [(b)] a sum of the $^4$He and
deuteron charge densities.
\end{description}

Provided corresponding experimental data for both densities are
used, the calculations show that a reasonable agreement with the
data can be obtained when the weight of the folding density
contribution is about $97.5\div98.5\%$ and the weight of the
contribution from the
sum of both densities is about $2.5\div1.5\%$.\\

\item The scheme has only one free parameter ($c_1$ or $c_2$) with
a clear physical meaning, namely, it is the weight of the one of the
contributions to the density of $^6$Li.\\

\item The behavior of the charge form factor of $^6$Li for $0<q\lesssim
2.7\, fm^{-1}$ is determined mainly by the folding contribution of
$^4$He and the deuteron densities to the charge density of $^6$Li
(the weight of this contribution is about $97.5\div98.5\%$).\\

\item The shell-model $\alpha-d$ cluster density of
$^6$Li (i.e. the sum of $^4$He and the deuteron charge densities)
is important (though with a small weight of about $2.5\div1.5\%$)
in the central nuclear region and, correspondingly, it is
responsible for the values of the charge form factor of $^6$Li at
large values of $q$ ($q\gtrsim 3\, fm^{-1}$).\\

\item The calculated within the suggested scheme charge rms radius of
$^6$Li agrees with the experimental estimations of this quantity.\\

\item We would like to pay attention to the following
facts:
\begin{description}
\item [(a)] the minimum of the experimental charge form factor of
the deuteron is at $q\approx4.2 fm^{-1}$ \cite{19,20}, \item
[(b)]the minimum of the experimental charge form factor of $^4$He
is at $q\approx3.2 fm^{-1}$ \cite{18}, \item [(c)] the minimum of
the experimental charge form factor of $^6$Li is at $q\approx
2.9\, fm^{-1}$ \cite{18,24,25,26,29,30}. Based on points {(a)} and
{(b)}, our estimations show that the latter minimum is determined
mainly by the contribution of the charge density and the
corresponding form factor of $^4$He.
\end{description}
\end{itemize}

\section*{Acknowledgments}

We would like to thank Professor Stephane Platchkov and Professor
Pedro Sarriguren for the valuable discussion. This work was partly
supported by the Bulgarian National Science Fund under the Contracts
No. $\Phi$-1416 and $\Phi$-1501.


\begin{thebibliography}{99}
\bibitem{1}
D. M. Dennison, {\it Phys. Rev.} {\bf 57} (1940) 454; {\it Phys.
Rev.} {\bf 96} (1954) 378; D. R. Inglis, {\it Rev. Mod. Phys.}
{\bf 25} (1953) 390; S. L. Kameny, {\it Phys. Rev.} {\bf 103}
(1956) 358; A. E. Glassgold, A. Galonsky, {\it Phys. Rev.} {\bf
103} (1956) 701.

\bibitem{2}
E. V. Inopin, B. I. Tishchenko, {\it J. Exp. Theor.
Phys.(Russian)} {\bf 37} (1959) 1309; {\bf 38} (1960) 1160; W.
Wadia, E. V. Inopin and M. Yusef, {\it J. Exp. Theor.
Phys.(Russian)}  {\bf 45} (1963) 1164; E. V. Inopin, A. A.
Kresnin, B. I. Tishchenko, {\it Sov. J. Nucl. Phys.(Russian)} {\bf
2} (1965) 802.

\bibitem{3}
M. Bouten, {\it Nuovo Cimento} {\bf 26} (1962) 63.

\bibitem{4}
Y. Yamaguchi, {\it Phys. Rev.} {\bf 95} (1954) 1628; D. R.
Harrington, {\it Phys. Rev.} {\bf 147} (1966) 685; H. Hebach, P.
Henneberg, {\it Z. Phys.} {\bf 216} (1968) 204; G. F. Bertsch,
W.Bertozzi, {\it Nucl. Phys.} {\bf A165} (1971) 199.

\bibitem{5}
H. Margenau, {\it Phys. Rev.} {\bf 59} (1941) 37; D. M. Brink,
``The Alpha-particle Model in Light Nuclei'', Int. School of
Physics ``E. Fermi'', {\bf course 36} (1965); D. M. Brink, H.
Friedrich, A. Weiguny, C. W. Wong, {\it Phys. Lett.} {\bf B33}
(1970) 143.

\bibitem{6}
T. Neff, H. Feldmeier and R. Roth, in {\it Proceedings of 21st
Winter Workshop on Nuclear Dynamics}, Breckenridge, Colorado, USA,
February 5-12, 2005.

\bibitem{7}
I. S. Gul'karov, {\it Issledovaniya yader electronami} (Atomizdat,
 Moscow, 1977).

\bibitem{8}
Yu. A. Kudeyarov, I. V. Kurdyumov, V. G. Neudatchin and Yu. F.
Smirnov, {\it Nucl. Phys.} {\bf A126} (1969) 36.

\bibitem{9}
T. I. Kopaleyshvili, I. Z. Machabeli, {\it Sov. J. Nucl. Phys.
(Russian)} {\bf 4} (1966) 37.

\bibitem{10}
Yu. A. Kudeyarov, Yu. F. Smirnov and M. A. Chebotarev, {\it Sov.
J. Nucl. Phys. (Russian)} {\bf 4} (1966) 1048.

\bibitem{11}
Yu. A. Kudeyarov, V. G. Neudatchin, S. G. Serebryakov and Yu. F.
Smirnov, {\it Sov. J. Nucl. Phys. (Russian)} {\bf 6} (1967) 1203.

\bibitem{12}
V. G. Neudatchin, Yu. F. Smirnov, {\it Nuclonnye associacii v
legkih yadrah} (Nauka, Moskow, 1969).

\bibitem{13}
G. Burleson, R. Hofstadter, {\it Phys. Rev.} {\bf 112} (1958)
1282.

\bibitem{14}
Yu. A. Kudeyarov, V. G. Neudatchin and Yu. F. Smirnov, {\it Izv.
Akad. Nauk USSR, Ser. Fiz.} {\bf 30} (1966) 235.

\bibitem{15}
Yu. A. Kudeyarov, Z. Matthiz, V. G. Neidachin and Yu. F. Smirnov,
{\it Nucl. Phys.} {\bf 65} (1965) 529.

\bibitem{16}
E. W. Schmid, K. Wildermuth and Y. C. Tang, {\it Phys. Lett.} {\bf
7} (1963) 263.

\bibitem{17}
T. de Forest, Jr., J. D. Walecka, {\it Adv. Phys.} {\bf 15} (1966)
1.

\bibitem{18}
V. V. Burov, D. N. Kadrev, V. K. Lukyanov and Yu. S. Pol', {\it
Phys. of Atom. Nucl.} {\bf Vol. 61}, no. 4 (1998) 525.

\bibitem{19}
D. Abbott {\it et al.}, {\it Eur. Phys. J.} {\bf A7} (2000) 421.

\bibitem{20}
D. Abbott {\it et al.}, {\it Phys. Rev. Lett.} {\bf 84} (2000)
5053.

\bibitem{21}
E. Tomasi-Gustafsson, G. I. Gakh and C. Adamu\v{s}\v{c}\'{i}n,
{\it Phys. Rev.} {\bf C73} (2006) 045204.

\bibitem{22}
S. Karataglidis, B. A. Brown, K. Amos and P. J. Dortmans, {\it
Phys. Rev.} {\bf C55} (1997) 2826; S. Karataglidis, P. J.
Dortmans, K. Amos and C. Bennhold, {\it Phys. Rev.} {\bf C61}
(2000) 024319.

\bibitem{23}
A. N. Antonov, D. N. Kadrev, M. K. Gaidarov, E. Moya de Guerra, P.
Sarriguren, J. M. Udias, V. K. Lukyanov, E. V. Zemlyanaya, G. Z.
Krumova, {\it Phys. Rev.} {\bf C72} (2005) 044307.

\bibitem{24}
V. V. Burov, V. K. Lukyanov, {\it Preprint JINR, Dubna} (1977)
{\bf R4-11098}.

\bibitem{25}
H. De Vries, C. W. De Jager and C. De Vries, {\it At. Data Nucl.
Data Tables} {\bf 36} (1987) 495.

\bibitem{26}
J. D. Patterson and R. J. Peterson, {\it Nucl. Phys.} {\bf A717}
(2003) 235.

\bibitem{27}
J. S. McCarthy, I. Sick and R. R. Whitney, {\it Phys. Rev.} {\bf
C15} (1977) 1396.

\bibitem{28}
C. R. Ottermann {\it et al.}, {\it Nucl. Phys.} {\bf A436} (1985)
688.

\bibitem{29}
L. R. Suelzle, M. R. Yearian and Hall Crannell, {\it Phys. Rev.}
{\bf 162} (1967) 992.

\bibitem{30}
G. C. Li, I. Sick, R. R. Whitney and M. R. Yearian, {\it Nucl.
Phys.} {\bf A162} (1971) 583.

\bibitem{31}
E. Friedman, A. Gal and J. Mares, {\it Nucl. Phys.} {\bf A579}
(1994) 518.

\end{thebibliography}
\end{document}